\documentclass[twocolumn,prl,prb,superscriptaddress]{revtex4-1}
\usepackage[latin9]{luainputenc}
\usepackage{graphicx}

\makeatletter

\providecommand{\tabularnewline}{\\}

\usepackage{pifont}
\usepackage{graphicx}
\usepackage{dcolumn}
\usepackage{bm}

\usepackage{hyperref}
\usepackage[caption=false,font=footnotesize]{subfig}

\usepackage{amsfonts}

\makeatother

\begin{document}

\title{Experimental demonstration of particle acceleration with normal conducting
accelerating structure at cryogenic temperature}

\author{Mamdouh Nasr}

\thanks{Corresponding author: mamdouh@slac.stanford.edu}

\affiliation{SLAC National Accelerator Laboratory, 94025 California, USA}

\author{Emilio Nanni}

\affiliation{SLAC National Accelerator Laboratory, 94025 California, USA}

\author{Martin Breidenbach}

\affiliation{SLAC National Accelerator Laboratory, 94025 California, USA}

\author{Stephen Weathersby}

\affiliation{SLAC National Accelerator Laboratory, 94025 California, USA}

\author{Marco Oriunno}

\affiliation{SLAC National Accelerator Laboratory, 94025 California, USA}

\author{Sami Tantawi}

\affiliation{SLAC National Accelerator Laboratory, 94025 California, USA}
\begin{abstract}
\textbf{Abstract:} Reducing the operating temperature of normal conducting
particle accelerators substantially increases their efficiency. Low-temperature
operation increases the yield strength of the accelerator material
and reduces surface resistance, hence a great reduction in cyclic
fatigue could be achieved resulting in a large reduction in breakdown
rates compared to room-temperature operation. Furthermore, temperature
reduction increases the intrinsic quality factor of the accelerating
cavities, and consequently, the shunt impedance leading to increased
system efficiency and beam loading capabilities. In this paper, we
present an experimental demonstration of the high-gradient operation
of an X-band, 11.424 GHz, 20-cells linear accelerator (linac) operating
at a liquid nitrogen temperature of 77 K. The tested linac was previously
processed and tested at room temperature. We verified the enhanced
accelerating parameters of the tested accelerator at cryogenic temperature
using different measurements including electron beam acceleration
up to a gradient of 150 MV/m, corresponding to a peak surface electric
field of 375 MV/m. We also measured the breakdown rates in the tested
structure showing a reduction of two orders of magnitude, $\times100$,
compared to their values at room temperature for the same accelerating
gradient. 
\end{abstract}
\maketitle

The operation of particle accelerators at high-gradient levels is
essential in future discovery machines for particle colliders and
free-electron lasers as well as advanced medical applications \cite{ILC_report,clic2018_report,FEL_mcneil2010x,VHEE_kokurewicz2019focused}.
The accelerating gradient is defined as the energy gained by a charged
particle moving along the axis of the accelerating structure per unit
length, in eV/m. Normal conducting (NC) accelerating structures have
been a leading candidate in high-gradient acceleration, with an accelerating
gradient above 100 MV/m, for their capability of sustaining higher
surface fields compared to their superconducting counterparts \cite{solyak2009gradient}.
The demanding applications for high-gradient operation and more compact
machines derive a continuous international effort to push the boundaries
of particle accelerator technology by providing compact accelerating
structures with optimized geometries and enhanced accelerating properties. 

A limiting factor in the operation of high-gradient accelerating structures
is the rf breakdowns on the surface which cause vacuum arcs that can
perturb acceleration \cite{Arcs_PhysRevB.81.184109}. Consequently,
a critical parameter in defining the operation of high-gradient accelerators
is the rf breakdown rates at the accelerating gradient of interest.
Studies have shown that breakdown rates are largely correlated to
the peak electric and magnetic fields on the surface and might be
explained by the movement of crystal defects due to the surface stress
from peak pulsed heating as well as the high electric field on the
inner accelerator surface \cite{Sc_grudiev2009new,nordlund2012defect,pulsedheating_dolgashev2003high,Pulsedheating_laurent2011experimental}.
There has been an effort to minimize the breakdown rates in high-gradient
accelerators by building accelerator structures from harder materials,
which showed lower breakdown rates compared to the ones built from
softer materials \cite{dolgashev2012progress,dolgashev2015high}.

Another approach that was experimentally investigated to enhance the
operation of NC accelerators is by operation at cryogenic temperatures
which substantially reduces the surface-resistance compared to room-temperature
operation. This reduction increases the shunt impedance and intrinsic
quality factor of the accelerating cavities leading to increased system
efficiency and beam loading capabilities; The shunt impedance is defined
as the square of the accelerating gradient divided by the power loss
per unit length of the accelerating structure, in $\Omega$/m. Moreover,
low-temperature operation increases surface hardness \cite{Hardness_reed1967low}
which is expected to largely reduce the breakdown rates compared to
room-temperature operation for the same accelerating gradient.

Few experiments investigated the operation of NC accelerating cavities
at cryogenic temperatures. In \cite{varian_mceuen1985high}, the operation
of accelerating cavities at 3 GHz and liquid nitrogen (LN) temperature
of 77 K was investigated. The experiment showed excessive deterioration
in the intrinsic quality factor for surface electric fields higher
than 150 MV/m. The authors concluded that this degradation is caused
by the high rf magnetic field on the surface. Other studies were performed
at low-gradient levels, below 50 MV/m \cite{02russia_saversky1993measurement,03KEK_iino2016high}.
These experiments verified the intrinsic quality factor of the tested
cavities at cryogenic temperatures. However, they did not demonstrate
high-gradient operation or study the breakdown rates in the tested
structures. An investigation performed by CERN studied the breakdown
rates for a set of cavities with frequencies between 21 and 35 GHz
at operating temperatures between 100 and 800 K \cite{CERN_braun2003frequency}.
The study showed no dependence of the breakdown rates on the temperature
of operation. 

A recent experimental effort aimed to provide a more elaborate picture
of the operation of high-gradient accelerating cavities at cryogenic
temperature \cite{cahill2018high}. The authors performed an experimental
investigation on a single-cell accelerating structure at 11.424 GHz
and 45 K. They reported reduced breakdown rates at cryogenic temperatures
compared to room temperature operation. This reduction enabled reaching
an accelerating gradient of 250 MV/m. They also reported a time-dependent
degradation in the intrinsic quality factor of the accelerating cavity
for accelerating gradients higher than 150 MV/m \cite{cahill2018rf}.
The experiments observed the onset of breakdowns at an accelerating
gradient of 250 MV/m, but were unable to characterize the breakdown
probability's dependance on accelerating gradient with the collected
statistics. 

The experimental effort in \cite{varian_mceuen1985high,02russia_saversky1993measurement,03KEK_iino2016high,cahill2018high,cahill2018rf}
showed a large promise in the operation of NC accelerators at cryogenic
temperatures for enhanced shunt impedance and reduced breakdown rates.
This effort was performed, however, for single-cell testing and/or
low-gradient operation, below 50 MV/m. Also, the lack of defined slopes
for the breakdown rates versus the accelerating gradient at cryogenic
temperature does not provide a clear understanding of the statistical
behavior of breakdown rates. Most importantly, these experiments did
not attempt to demonstrate particle acceleration with the tested structures.
To the best of the authors' knowledge, there has never been an experimental
demonstration of the operation of multi-cell NC accelerating structures
at cryogenic temperature and high-gradient levels.

In our work, we develop a practical approach and methodology for the
high-gradient operation of multi-cell NC accelerating structures at
cryogenic temperature. This development is a necessary milestone in
the road of adapting Cryogenic-NC-accelerators in many practical applications
including future discovery machines. Practical and economic aspects
are thus important. We decided to operate at a cryogenic temperature
of LN of 77 K. Operating with the LN cooling system results in much-reduced
system cost compared to the complicated setup required for the operation
with LHe, and still provides much-enhanced accelerating parameters
compared to room temperature operation. We verified the accelerating
parameters using many approaches including energy-gain measurements
with electron-beam. We also performed an extensive study on the breakdown
rates for the same accelerator structure at room and cryogenic temperatures. 

\begin{figure}
\begin{centering}
\includegraphics[width=0.4\textwidth]{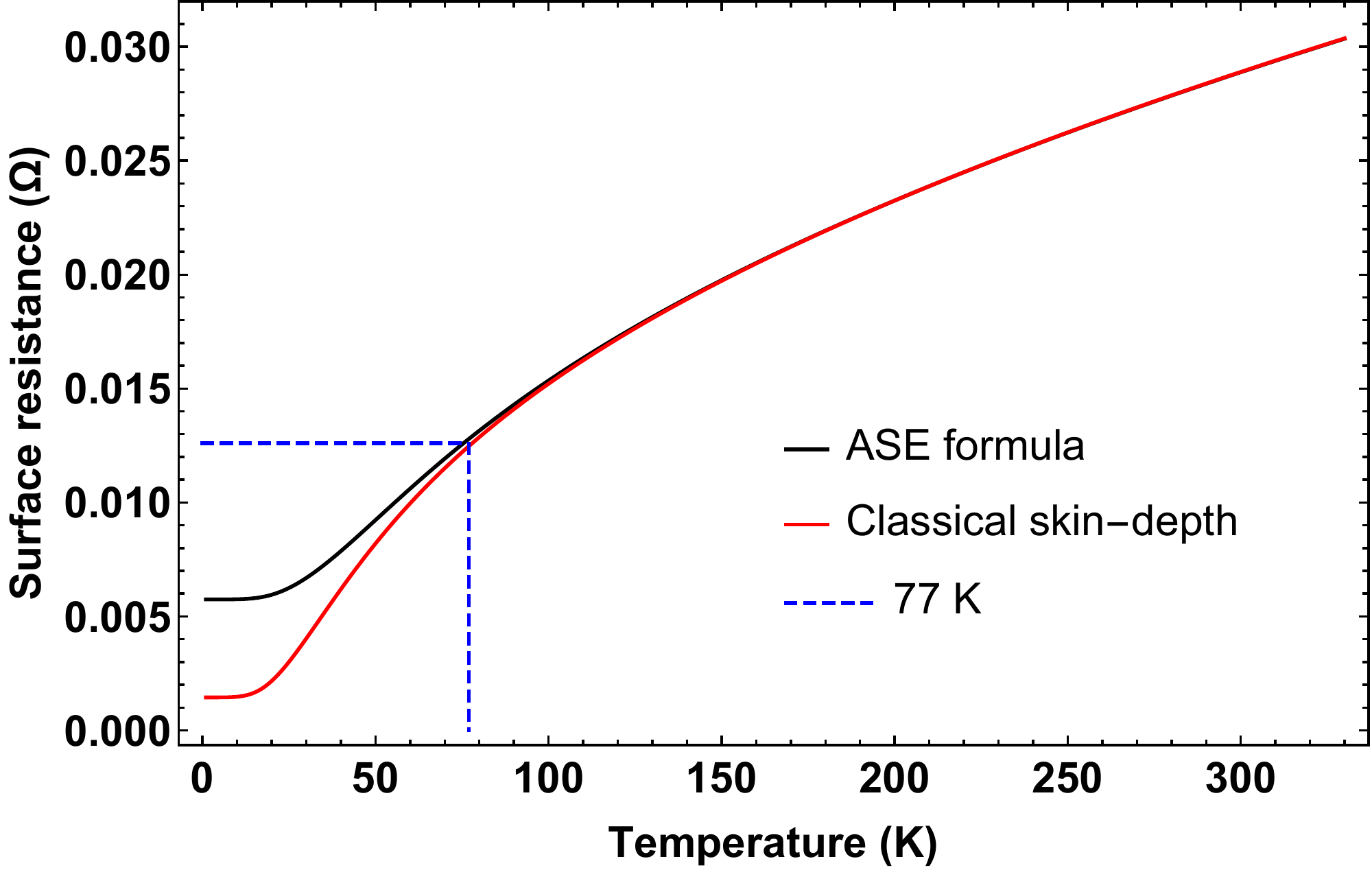}
\par\end{centering}
\caption{\label{fig:zs_vs_temp} At low temperatures the surface resistance,
predicted by ASE, saturates to higher values than the ones calculated
from the classical skin depth. Operating at 77 K results in a reduction
in the surface resistance of a factor of 2.25 compared to 300 K. }
\end{figure}

The rf surface resistance of good-conductors at cryogenic temperatures
is well studied and explained using the theory of anomalous skin effect
(ASE) \cite{ASE_reuter1948theory}. At low temperatures and high frequencies,
the mean free path of electrons becomes comparable or larger than
the classical skin depth for good conductors. In this case, the electrons
that contribute the most to the current are the ones that spend their
entire mean free path in the skin depth, and the surface resistance
saturates to a higher value than the one predicted by the classical
skin depth. Fig. \ref{fig:zs_vs_temp} shows the surface resistance
for copper (RRR=400) versus temperature at 11.424 GHz. In our calculations,
we used the Bloch-Grüneisen Formula in \cite{Bulk_cond_matula1979electrical}
to calculate the bulk conductivity versus temperature, and the equations
for the ASE in the simplified form presented in \cite{Emma_stupakov2015resistive}.
At low temperatures, the surface resistance, predicted by the ASE,
saturates to higher values than the ones calculated from the classical
skin depth. Operating at 77 K results in a reduction in the surface
resistance of a factor of 2.25 compared to 300 K. 

\begin{figure*}
\begin{centering}
\includegraphics[width=0.88\textwidth]{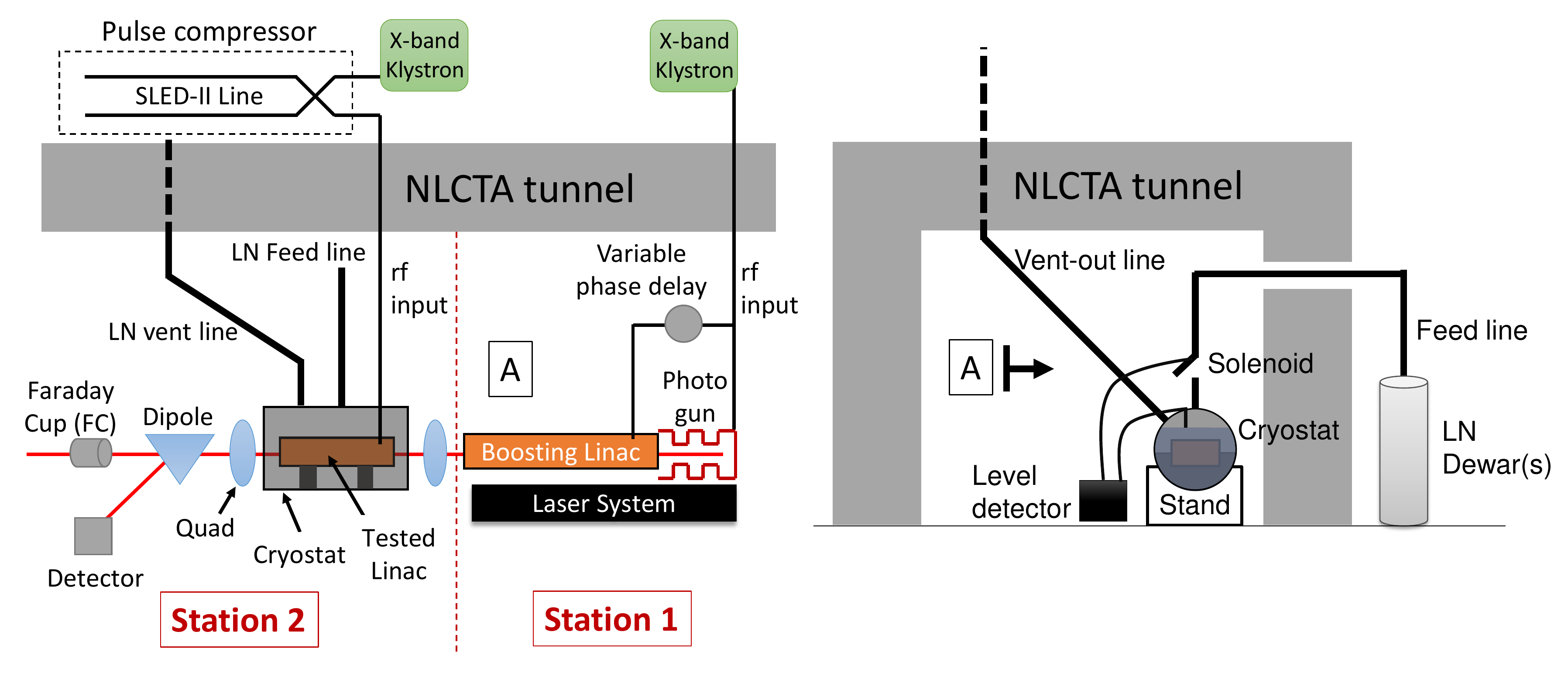}
\par\end{centering}
\begin{centering}
(a)\hspace*{8cm}(b)
\par\end{centering}
\caption{\label{fig:experiment_setup} (a) The experimental setup for accelerator
testing at XTA facility at SLAC. XTA is divided into two stations,
referenced as station 1 and station 2. The first station generates
the electron beam used in the experiment, and the tested accelerator
is installed in the second station. (b) A foam-insulated stainless-steel
line delivers LN from Dewars outside the tunnel to the cryostat where
the tested linac is setting inside, through a hole in the tunnel wall.
A level detector is inserted inside the cryostat and provides a control
signal to a solenoid that controls the flow of the LN in the feed
line. A vent-out line transport the boiled LN outside the tunnel. }
\end{figure*}

We tested an X-band, 11.424 GHz, 20-cells, standing-wave linear accelerator
(linac) structure which has been previously processed and tested at
room temperature up to a gradient of 140 MV/m \cite{Dist_tantawi2020design}.
Table \ref{tab:acc_parameters} shows the tested linac parameters
at 300 K and 77 K. We modified the existing setup at X-band Test Accelerator
(XTA) station at the NLCTA facility at SLAC to accommodate high-power
experiments at cryogenic temperatures. The experimental setup is shown
in Fig. \ref{fig:experiment_setup}(a). XTA is divided into two stations,
referenced in Fig. \ref{fig:experiment_setup}(a) as stations 1 and
2; Each station is fed with an independent rf source. The first station
is used to generate the electron beam from an X-band photoinjector
and then accelerate the beam to about 46 MeV, in our experiment, using
an NLC style X-band accelerator \cite{wang2004accelerator}. The second
station feeds the tested linac, which is installed after the station
1 linac. An X-band klystron feeds the accelerator under testing with
rf power. The peak rf power of the klystron is 40 MW, and we installed
a pulse compressor (multimoded SLED-II \cite{SLED_II_tantawi2005high})
after the klystron to push the rf power above the klystron limit.
Quadrupoles are used for beam focusing and trajectory adjustment,
a spectrometer is placed after the distributed-coupling linac to measure
the beam energy, and a Faraday cup (FC) is installed downstream for
charge measurements. We placed a dark-current detector downstream
from the tested accelerator to detect dark-current spikes and count
the breakdown rates in our tested accelerator.

\begin{table}
\caption{\label{tab:acc_parameters}Summary of the accelerating parameters
of the tested accelerating structure at 300 and 77 K. The peak fields
are calculated for average accelerating gradient of 100 MV/m.}

\begin{ruledtabular} %
\begin{tabular}{lll}
Parameter & 300 K & 77 K\tabularnewline
\hline 
Frequency (GHz) & 11.402 & 11.438\tabularnewline
$Q_{0}$ & 10000 & 22500\tabularnewline
$Q_{ext}$ & 10000 & 10000\tabularnewline
Shunt impedance (M$\Omega$/m) & 155  & 349\tabularnewline
Peak surface E (MV/m) & 250  & 250\tabularnewline
Peak surface H (MA/m) & 0.575  & 0.575 \tabularnewline
Iris diameter ( mm) & 2.6 & 2.6\tabularnewline
Length (cm) & $26$ & $26$\tabularnewline
\end{tabular}\end{ruledtabular}
\end{table}

Figure \ref{fig:experiment_setup}(b) illustrates the cooling system
used for our cryogenic experiment. We drilled a hole through the tunnel
wall, perpendicular to the beam axis. A foam-insulated stainless-steel
line delivers LN from Dewars outside the tunnel to the cryostat, where
the tested linac is sitting inside. A level detector is inserted inside
the cryostat and provides a control signal to a solenoid that controls
the flow of the LN in the feed line. A vent-out line transports the
boiled LN outside the tunnel. The used cryostat is a simple cylindrical
stainless-steel container with insulation foam from the outside. The
linac structure is sitting inside the cryostat on two half-circular
supports with a rectangular opening for the structure to set inside.
The cryostat and the linac share the same beam axis and there are
two transport tubes under-vacuum from both sides of the linac to the
cryostat walls perpendicular to the beamline. The cryostat has an
opening for the feed lines of the LN and the input rf power, and small
openings for the LN level detector and vent-out line. 

\begin{figure}
\begin{centering}
\includegraphics[width=0.4\textwidth]{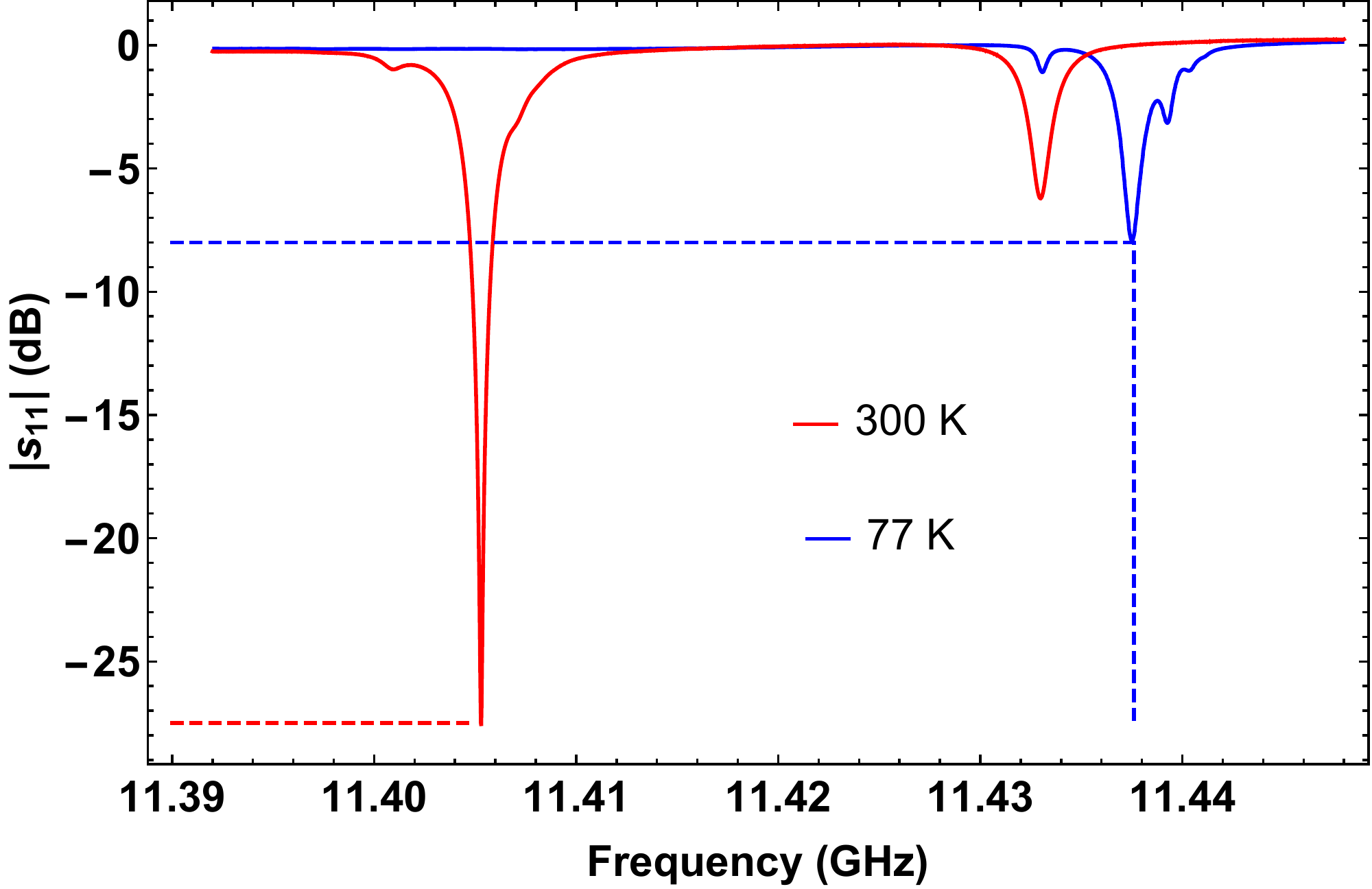}
\par\end{centering}
\caption{\label{fig:s11_vs_temp} The reflection coefficient at the rf input
to the accelerator at 300 K and 77 K. As the accelerator structure
cools down, it shrinks leading to a shift of the resonance frequency
of 36 MHz. The tested linac is designed for critical-coupling at 300
K, $\left|s_{11}\right|=-27$ dB at resonance, and becomes over-coupled
at 77 K, $\left|s_{11}\right|=-8$ dB at resonance, because of the
reduced surface resistance which increases the intrinsic quality factor
of the accelerating cavity while the external one remains constant.}
\end{figure}

Before installing the cryostat in the beamline, we performed a cold-test
of our accelerator structure at the cryogenic temperature of LN. We
installed the accelerator structure, under-vacuum, inside the cryostat,
and used LN Dewar to supply LN directly into the cryostat. A VNA was
connected to the rf input of the structure, through a vacuum window,
to measure the reflection coefficient at the rf input port to the
accelerator. Figure \ref{fig:s11_vs_temp} shows the reflection coefficient
at the rf input to the accelerator at 300 K and 77 K. As the accelerator
structure cools down, it shrinks, and thus the resonance frequency
shifts to a higher value. We measured a resonance shift of 36 MHz
from 300 K to 77 K which is in very good agreement with the calculated
value using the thermal expansion of copper \cite{nix1941thermal_expansion}.
The distributed-coupling linac is designed for critical-coupling at
300 K and becomes over-coupled at 77 K because of the reduced surface
resistance which increases the intrinsic quality factor, $Q_{0}$,
of the accelerating cavity. Note that the external quality factor
remains constant because of the minimal effect of the temperature
change on the power-coupling irises to the accelerating cells. We
should emphasize that the increase of the coupling coefficient, $\beta=Q_{0}/Q_{e}$,
for the accelerating structure from 1 to 2.25 from 300 to 77 K represents
a large increase in the beam-loading capability for the same accelerator
structure just by cooling it down to cryogenic temperature.

\begin{figure}
\begin{centering}
\includegraphics[width=0.4\textwidth]{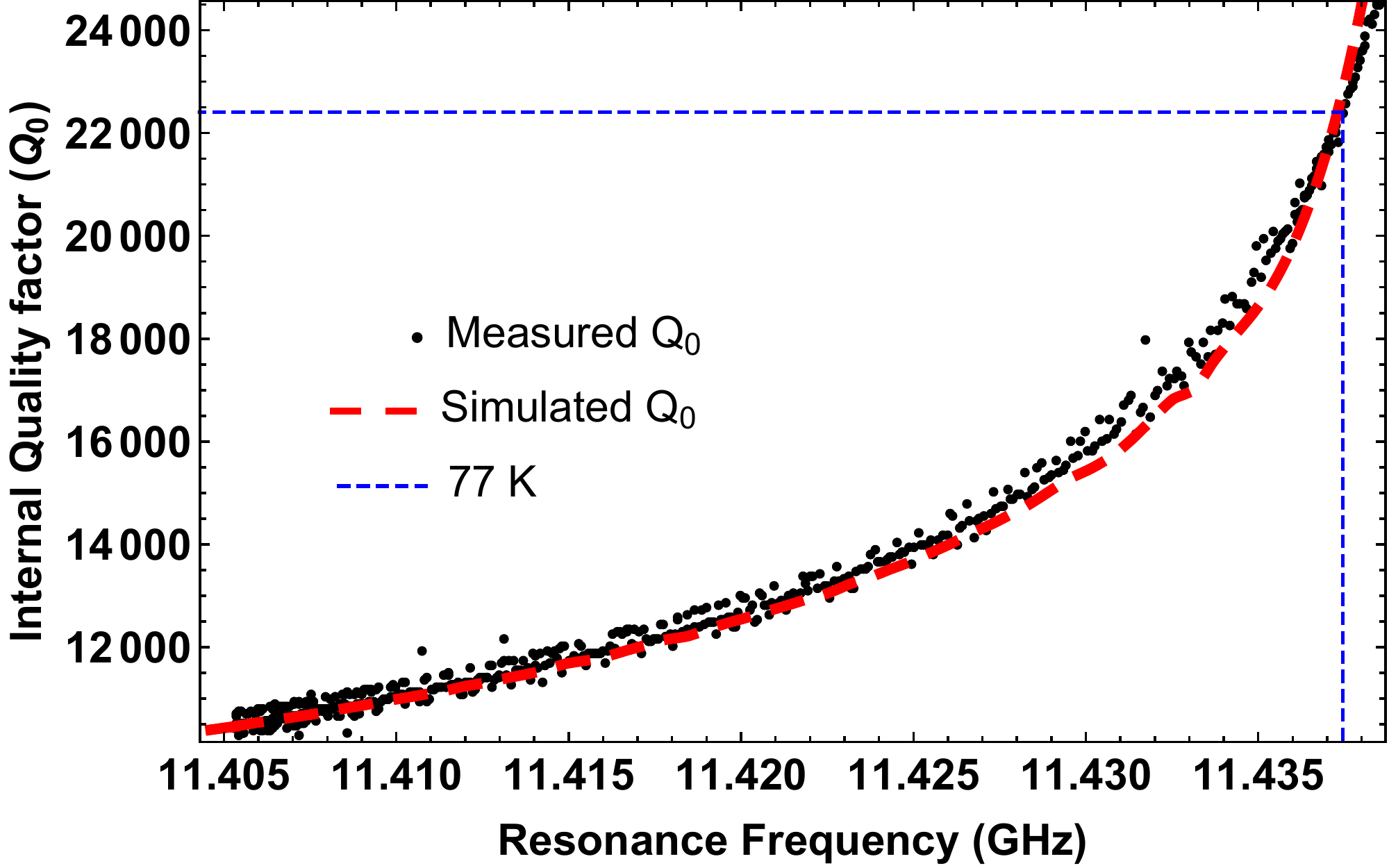}
\par\end{centering}
\caption{\label{fig:Q_vs_temp} The intrinsic quality factor of the accelerating
structure versus the resonance frequency. Black dots and dashed-line
are the measured and simulated values, respectively. The measured
values are obtained using Q-circle fitting of the measured reflection
coefficient at the rf input port. In our simulations, the temperature
at each point is used to calculate the surface resistance using the
theory of ASE and simulate the intrinsic quality factor.}
\end{figure}

We did another test to measure the change in the quality factor and
resonance frequency for the accelerator structure versus temperature.
We supplied LN into the cryostat until the structure temperature stabilized.
We then stopped the LN supply and measured the reflection coefficient
as the structure warms-up to room temperature. Figure \ref{fig:Q_vs_temp}
shows the measured intrinsic quality factor of the accelerating structure
versus resonance frequency from 300 to 67 K. The measured values of
the intrinsic quality factor are obtained using Q-circle fitting of
the measured reflection coefficient at the rf input port \cite{Qcircle_kajfez1994linear}.
We compared the measured values of the intrinsic quality with our
simulations, represented by the dashed line in Fig. \ref{fig:Q_vs_temp}.
In our simulations, we calculated the temperature at each point from
the measured shift of the resonance frequency, from its value at room
temperature, and the thermal expansion coefficient of copper \cite{nix1941thermal_expansion}.
We then used the temperature value to calculate the surface resistance
using the theory of ASE, previously presented in Fig. \ref{fig:zs_vs_temp}.
The obtained surface resistance is used to simulate the intrinsic
quality factor of the accelerator structure. The results show very
good agreement between the measurement and simulation. The lowest
measured temperature is 67 K, however, for high-power operation, the
temperature stabilizes at 77 K with an intrinsic quality factor of
22500. 

\begin{figure}
\begin{centering}
\includegraphics[width=0.38\textwidth]{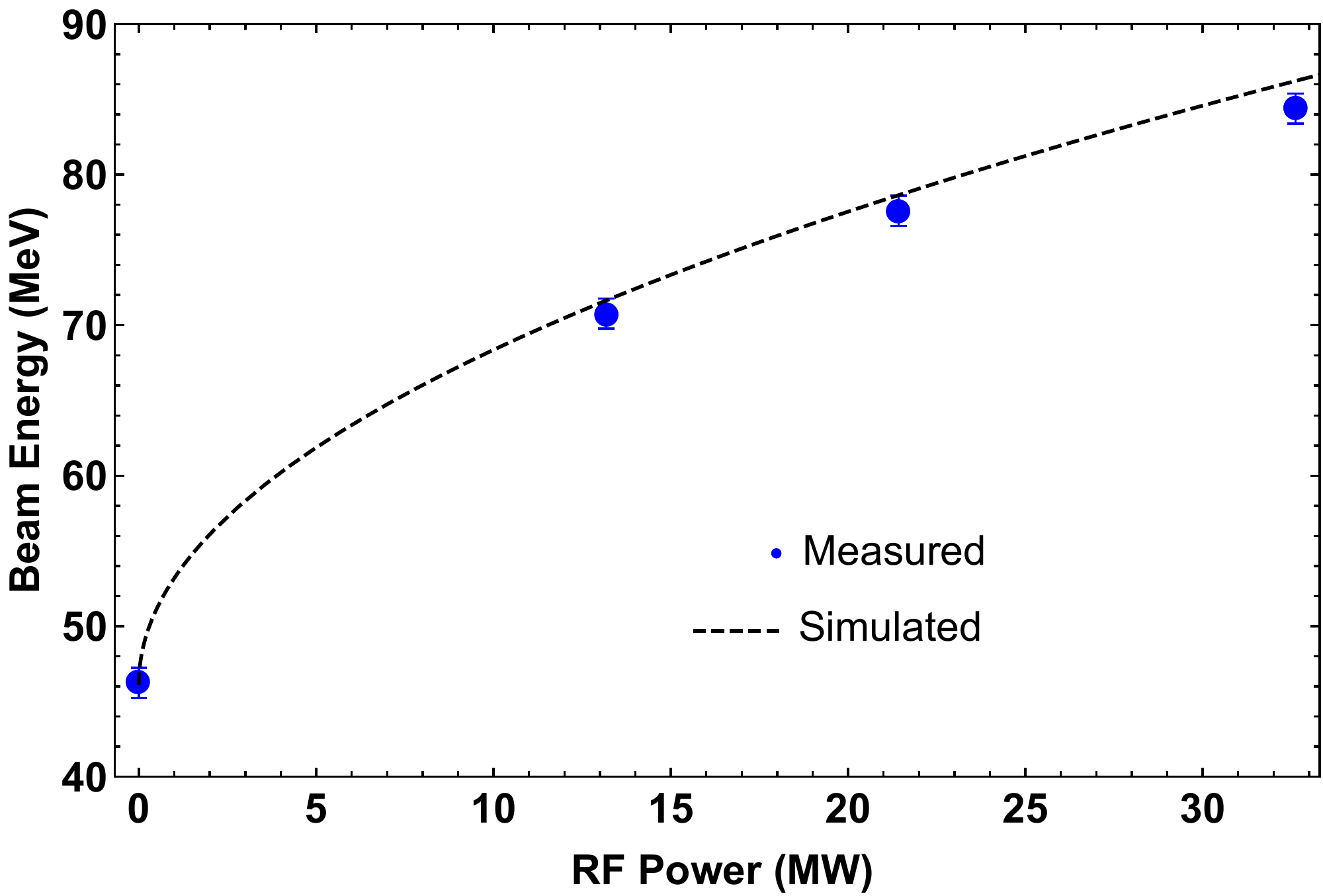}
\par\end{centering}
\caption{\label{fig:energy_gain} The measured energy gain of an electron beam
moving down the axis of the tested linac for a compressed 200 ns square
rf pulse with input power of 13-33 MW achieving accelerating gradient
from of 100-150 MV/m, and peak surface electric field of 250-375 MV/m.
Blue dots and dashed-line are the measured and simulated values, respectively.
Simulations used the measured input rf pulse to the structure and
the cavity model to calculate the energy gain at 77 K.}
\end{figure}

After verifying the quality factor of our accelerating structure at
low-level rf, we installed the cryostat in the XTA beamline. The goal
of our experiment is to verify the accelerating parameters of our
structure and to collect breakdown statistics for high-power operation
at 77 K. We measured the energy gain of an electron beam moving down
the axis of the tested linac and compared the results with our simulations
using the measured input pulse to the structure and the cavity model,
substituting with the accelerating parameters from Table \ref{tab:acc_parameters}
at 77 K \cite{wangler2008rf}. We performed a set of measurements
using a compressed 200 ns square rf pulse with an input power of 13-33
MW achieving an accelerating gradient from 100-150 MV/m, and a peak
surface electric field of 250-375 MV/m. Figure \ref{fig:energy_gain},
shows the measured energy gain versus input rf power showing very
good agreement between the measurements and simulations, less than
4\% deviation. 

\begin{figure}
\begin{centering}
\includegraphics[width=0.4\textwidth]{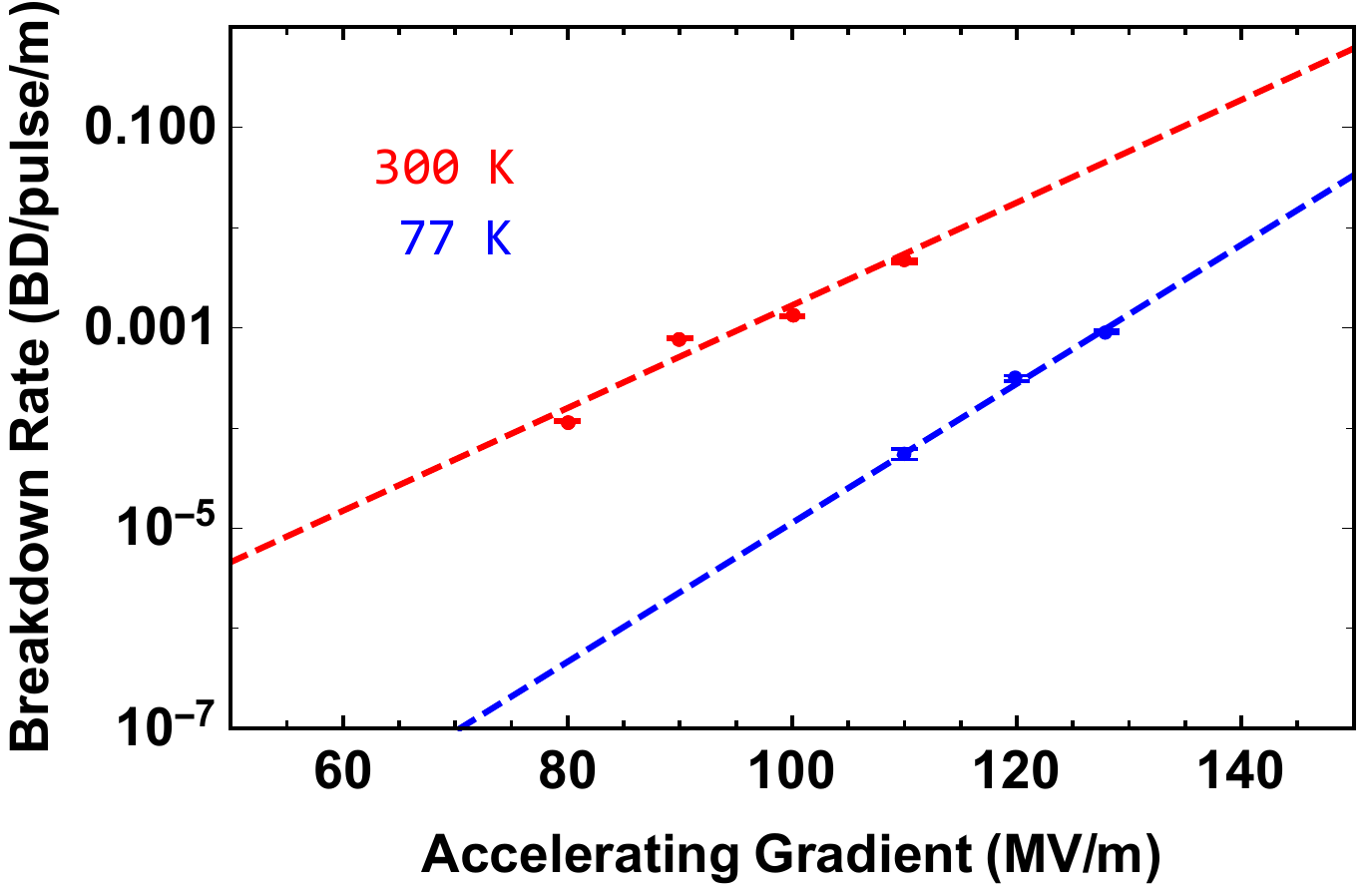}
\par\end{centering}
\caption{\label{fig:BDR} The collected breakdown rates, with the fitted slope,
of the tested accelerator structure at 300 and 77 K. The breakdown
rates at 300 K are obtained from our previous testing of the same
accelerator structure at room temperature \cite{Dist_tantawi2020design}.
The results show a reduction of two orders of magnitude, $\times100$,
in breakdown rates at the same gradient levels from 300 to 77 K operation. }
\end{figure}

Figure. \ref{fig:BDR} shows the collected breakdown rates, with the
fitted slope, for the tested accelerator structure at 77 K with an
accelerating gradient of 110-130 MV/m. The breakdown rates are compared
with the results from our previous testing of the same accelerator
structure at 300 K. For both operating temperatures, the data was
collected for a 400 ns stepped pulse with a flat gradient of 200 ns.
The results show a reduction of two orders of magnitude, $\times100$,
in breakdown rates at the same gradient levels from 300 to 77 K operation.
We believe that this reduction in breakdown rates is correlated to
the increased hardness of copper at cryogenic temperature. This reduction
and the observed increase in the slopes of the breakdown lines at
77 K is in agreement with the previously observed behavior for accelerating
cavities built with hard versus soft copper \cite{dolgashev2012progress,dolgashev2015high}.
The results are also in agreement with the reduced breakdown rates
reported in \cite{cahill2018high}. Our results provide comprehensive
breakdown statistics for the high-gradient operation of NC accelerators
at cryogenic temperature. These results give a more realistic intuition
on the dependence of breakdown rates on the operating temperature
compared to the study in \cite{CERN_braun2003frequency} that showed
no temperature dependence of the breakdown rates in the tested cavities.
This reduction in breakdown rates at lower temperatures of operation
is also predicted using different proposed breakdown models for normal
conducting accelerating structures \cite{Sc_grudiev2009new,nordlund2012defect,pulsedheating_dolgashev2003high,Pulsedheating_laurent2011experimental}.

The presented experiment provides the first demonstration of high-gradient
acceleration of an electron-beam at a cryogenic temperature of 77
K; We reached a gradient level of 150 MV/m with a peak surface electric
field of 375 MV/m. The structure achieved a shunt impedance of 349
M$\Omega$/m and $\times2.25$ enhancement in the beam loading capabilities
compared to 300 K operation. The experiment provides a practical and
cost-effective approach for the high-gradient operation of NC accelerators
at cryogenic temperatures of 77 K with LN cooling. We presented a
comprehensive breakdown study of the tested accelerator structure
at cryogenic temperature and high-gradient operation showing two orders
of magnitude, $\times100$, reduction in breakdown rates from 300
to 77 K. The reduced breakdown rates agree with our understanding
of the correlation between reduced breakdown rates and the increased
material hardness. This experimental investigation provides a critical
milestone for the practical use of NC accelerating systems at cryogenic
temperatures which motivates many proposals for future discovery machines
with optimized cost and performance parameters \cite{C3_bane2018advanced,C3_rosenzweigc3,rosenzweig2020ultra}. 

\bibliographystyle{apsrev4-1}
\bibliography{PRLCryo}

\end{document}